\documentclass{appolb}

\usepackage{graphicx}
\usepackage{subfigure}
\usepackage{bm}

\makeatletter
\newcommand\erfc{\mathop{\operator@font erfc}\nolimits}
\def\slashchar#1{\setbox0=\hbox{$#1$}
   \dimen0=\wd0 \setbox1=\hbox{/} \dimen1=\wd1
   \ifdim\dimen0>\dimen1 \rlap{\hbox to \dimen0{\hfil/\hfil}} #1
   \else  \rlap{\hbox to \dimen1{\hfil$#1$\hfil}} / \fi}

\makeatother

\begin{document}
\bibliographystyle{h-elsevier3} 
\title{Dissipation in the very early stage of the hydrodynamic evolution
in relativistic heavy ion collisions}
\author{Piotr Bo\.zek
\footnote{email:~Piotr.Bozek@ifj.edu.pl}
\address{Institute of Physics, Rzesz\'ow University, PL-35959 Rzesz\'ow, Poland}
\and
  \address{The H. Niewodnicza\'nski Institute of Nuclear Physics,
PL-31342 Krak\'ow, Poland}}
%

\maketitle

\begin{abstract}
We propose a modification of the hydrodynamic model of the dynamics in
 ultrarelativistic
nuclear collisions.
A  modification of the energy-momentum tensor at the initial stage 
describes the lack of isotropization of the pressure. Subsequently, 
  the pressure is relaxing  towards the equilibrium isotropic form 
in the local comoving
 frame. 
Within  the Bjorken scaling solution a bound is found on the 
decay time of the initial anisotropy of the energy-momentum tensor.
For the strongest dissipative effect allowed,  we find a
 relative entropy increase  of about 30\%, a significant
 hardening of the transverse momentum spectra, and no effect on the HBT radii.
\end{abstract}



\section{Introduction\label{sec:intro}}

Relativistic hydrodynamics is a common framework for the modelling of the 
expansion and the freeze-out  of the hot and dense region formed in 
ultrarelativistic 
nuclear collisions \cite{Kolb:2003dz}.  Depending on the initial conditions
 and on the freeze-out temperature a 
substantial amount of collective flow can be  built up 
during the hydrodynamic evolution. 
The physical picture of the freeze-out 
combines a thermal emission of particles 
from the local fluid element with the collective flow due to the 
movement of the fluid element. Choosing  azimuthally 
asymmetric initial conditions for collisions at finite impact parameter,
the hydrodynamic evolution generates elliptic asymmetry in the momentum
 distributions
 similar to the
experimentally  observed one. 
Perfect fluid dynamics (without shear viscosity) 
generates a strong elliptic flow, and a relatively moderate radial flow. The 
evolution has to be followed for a long time in order to reproduce 
the transverse momentum spectra and the elliptic flow for different particle 
species  \cite{Kolb:2003dz}. It means that the freeze-out temperature is low. 
Stronger asymmetry of the shape of the initial interaction zone 
is predicted by the Color-Glass-Condensate model \cite{Drescher:2006ca}
 or  is due to
event-by-event eccentricity fluctuations \cite{Aguiar:2001ac}; naturally it 
 leads to stronger elliptic flow. 
The strong 
elliptic flow is reduced if shear viscosity is important for the 
liquid formed in ultrarelativistic collisions at RHIC energies 
\cite{Teaney:2003kp}. 
Shear viscosity reduces the work of the 
fluid in the longitudinal direction, which means
 a slower colling  rate, and consequently a larger 
pressure to drive the transverse radial flow. For the  Hanbury-Brown-Twiss
(HBT) radii the presence of viscous terms at the 
freeze-out could be important, but
 does not explain  the observed HBT 
radii \cite{Romatschke:2007jx}. 

The actual value and the range of applicability of the ideal fluid 
hydrodynamics 
in the development of the dense medium in the collision is uncertain.
Recent calculations indicate that the shear viscosity 
coefficient should be small \cite{Song:2007fn},
 of the order of  the conjectured lower bound \cite{Kovtun:2004de}.
The build up of the transverse collective flow is greatly facilitated by the 
shear viscosity but the effect is reduced by viscous 
corrections at the freeze-out;
 calculations indicate that a required amount 
of transverse flow  appears.

 Even if we assume
 that the bulk 
evolution of the hot and dense matter created in heavy ion collisions
 is well described by the ideal fluid hydrodynamics, 
dissipative effects may appear  in the 
very early stage of the collision. 
The physical motivation for this picture is the fact that
 in the initial stage, for evolution times before $1$fm/c, one cannot expect a 
full local thermalization of the medium. This effect can be phenomenologically 
taken into account in the hydrodynamic evolution as a modification 
of the initial condition for the energy-momentum tensor.
This idea is at the origin of  an extreme scenario in the collision dynamics,
 where the expansion is two-dimensional, only in the transverse direction
\cite{Heinz:2002rs,Bialas:2007gn,Chojnacki:2007fi}.  Assuming that 
the anisotropy of  local momentum distribution does not equilibrate
 during the evolution of the fireball, a strong effect on the transverse 
expansion is seen. The 
 transverse flow builds up faster \cite{Heinz:2002rs,Bialas:2007gn}. 

In the present work, we study a more realistic scenario
 where the anisotropy in
 the initial conditions dissipates with time and eventually the limit 
of the  ideal fluid with an isotropic pressure is recovered. 
The dynamics of the equilibration of the local pressure is based 
on the second order dissipative
 relativistic hydrodynamics with shear viscosity \cite{IS}.
In this paper we study of the effects of the initial anisotropy only, therefore 
 we assume 
that the shear viscosity coefficient is zero and the energy-momentum tensor 
relaxes to the ideal fluid one. A new phenomenological parameter is
introduced, the relaxation time of the pressure anisotropy. 
We study the modified hydrodynamic evolution with initial anisotropy
 for two different geometries of the flow~: the Bjorken scaling solution with
 longitudinal expansion only, and the boost-invariant flow in the 
longitudinal direction with an azimuthally symmetric expansion in the
 transverse directions.
The dissipation in the early stage of the evolution leads to an 
increase of the entropy of the system. This effect must be compensated by a 
suitable retuning of the initial conditions. As a result,  a slight 
hardening of the transverse momentum spectra of emitted 
particles is found  and almost no effect on  HBT radii in central
 collisions is visible. 

\section{Early dissipation}

Nonzero shear viscosity is believed to be the most important
 modification of 
 ideal fluid  hydrodynamics in ultrarelativistic collisions 
\cite{Teaney:2003kp,Muronga:2001zk}.
The energy-momentum tensor is modified by the shear tensor $\pi^{\mu\nu}$
\begin{equation}
T^{\mu\nu}=(\epsilon+p)u^\mu u^\nu -g^{\mu\nu}p+\pi^{\mu\nu} \ ,
\end{equation}
$u^\mu$ is the velocity of the fluid element, the energy density 
$\epsilon$ and the pressure $p$ are related by the equation of state.
Besides the equation of state and the hydrodynamic equations
\begin{equation}
\partial_\mu T^{\mu \nu}=0
\end{equation}
 one has  a relaxation equation for the shear tensor \cite{IS} 
\begin{eqnarray}
\tau_\pi\Delta^\mu_\alpha\Delta^\nu_\beta u^\gamma \partial_\gamma
\pi^{\alpha\beta}&=&\eta <\nabla^\mu u^\nu> -\pi^{\mu\nu} \nonumber \\
& & -\frac{1}{2} \eta T \pi^{\mu\nu} \partial_\alpha
\left(\frac{\tau_\pi u^\alpha}{\eta T}\right)
+ \tau_\pi \pi^{\alpha(\mu}\omega^{\nu) }_\alpha \ ,
\label{eq:relax1}
\end{eqnarray}
$\Delta^{\mu\nu}=g^{\mu\nu}-u^\mu u^\nu$.
The last term contains the vorticity of the fluid
 $\omega^{\mu\nu}=\Delta^{\mu\alpha}\Delta^{\nu\beta}
\left(\partial_\alpha u_\beta-\partial_\beta u_\alpha\right)$, which 
  is zero for the flows  considered here.
$T$ is the local temperature and
 \begin{equation}<\nabla^\mu u^\nu>=\nabla^\mu u^\nu+\nabla^\nu
 u^\mu-\frac{2}{3}\Delta^{\mu\nu}\nabla_\alpha u^\alpha \ \ ,
\end{equation}
$\eta$ is the shear viscosity coefficient, and $\tau_\pi$ is the 
relaxation time of the shear tensor.
The term $\eta<\nabla^\mu u^\nu>$ in Eq. (\ref{eq:relax1}) 
is the Navier-Stokes (first-order) 
viscous correction to the energy-momentum tensor.
The viscosity and the relaxation time are related to the rates of  
equilibration processes in the plasma. The viscosity coefficient 
 can be estimated to be 
$\eta\simeq 1.04 s$ for a Boltzmann massless gas \cite{Venugopalan:1992hy},
 $\eta\simeq 0.7 - 1.1 s$ 
for ($N_f=0$) QCD \cite{Arnold:2000dr}, and is expected \cite{Kovtun:2004de}
to fulfill the bound $\eta\ge \frac{1}{4\pi}s$, where $s$ is
 the entropy density.
Estimates for the ratio of $\tau_\pi$ and $\eta$   range from
$\tau_\pi/\eta=\frac{6}{T s}$ to $\tau_\pi/\eta \simeq \frac{0.2}{T s}$.

In the early stage of the collision the flow is dominated 
by the flow in the longitudinal ($z$) direction. We assume a 
Bjorken flow with the  four-velocity of the form $u^\mu=(t/\tau,0,0,z/\tau),
 \ \tau=\sqrt{t^2-z^2}$. For the Bjorken boost invariant
scaling solution of hydrodynamic equations  the stress tensor is
\begin{equation}
\pi^{\mu\nu}=\left( \begin{array}{cccc} -\sinh^2 y& 0 & 0 & -\sinh y \cosh y \\
0 & 1/2 & 0 & 0 \\
0 & 0 & 1/2 & 0 \\
-\sinh y \cosh y & 0 & 0 & -\cosh^2 y \end{array} \right) \Pi
\end{equation}
where $\Pi$ is the solution of a dynamical equation \cite{Muronga:2001zk}
\begin{equation}
\tau_\pi \frac{d \Pi(\tau)}{d\tau} = \frac{4}{3}\frac{\eta}{\tau}-\Pi(\tau)
-\frac{\Pi(\tau)}{2}\left( \frac{\tau_\pi}{\tau}
+\frac{T \eta}{\tau_\pi}\frac{d}{d\tau}
\left(\frac{\tau_\pi}{T\eta}\right)\right) 
\label{eq:dynamical}
\end{equation}
and $y$ is the rapidity of the fluid element.
In the first-order dissipative hydrodynamics we have
 the steady-flow result 
\begin{equation}
\Pi(\tau)=\frac{4\eta}{3\tau}
\label{eq:ns}
\end{equation}
 for the shear viscosity corrections to the energy-momentum tensor
\cite{Teaney:2003kp}. 
 Solving the 
 dynamical equation 
(\ref{eq:dynamical}) requires the knowledge of the 
value of  the viscous strength $\Pi(\tau_0)$ at the initial time.
 The role of the initial condition in the further evolution until the freeze-out
depends on the relaxation time $\tau_\pi$. 
For a typical choice of parameters $\tau_\pi$ and $\eta$, the initial
 value of $\Pi(\tau)$ relaxes fast and is not determinant for
 the further evolution \cite{Baier:2006gy,Dumitru:2007qr,Song:2007fn}.
The initial value for viscous corrections should fulfill the conditions of
 the applicability of hydrodynamics
 with viscosity $\sqrt{\pi^{\mu\nu}\pi_{\mu\nu}}\ll p$. 
In practice, for the Bjorken flow,
 the condition $\frac{\Pi(\tau_0)}{p+\epsilon}\le 1$ 
or $\Pi(\tau_0)\le p$ is used.
 Another natural restriction is the condition 
that the relative  viscous correction 
 $\frac{\Pi}{p+\epsilon}$ decreases with time \cite{Dumitru:2007qr}.

In this paper we study a different type 
of local deviation from equilibrium in the hydrodynamic evolution.
The initial local momentum distributions in 
the transverse and longitudinal directions could be 
 different, there is no reason
 to expect instantaneous
 isotropization of the momentum distributions. One could assume
 that this asymmetry remains until the 
freeze-out \cite{Heinz:2002rs,Bialas:2007gn}. Such a  hydrodynamic 
evolution in the transverse direction only gives stronger transverse 
flow, faster expansion, small HBT radii and a
realistic elliptic flow \cite{Heinz:2002rs,Bialas:2007gn,Chojnacki:2007fi}.
A fast build up of the transverse flow is 
compatible with experimental indications of a rapid 
break-up and hadronization of the fireball.
The assumption that the local momentum distribution is of the form 
\cite{Heinz:2002rs,Bialas:2007gn}
\begin{equation}
f(p)\propto \delta(y-\eta) f_{thermal}(p_T) 
\end{equation}
where $y$ and $\eta$ are the rapidity and the space-time
 rapidity, whereas
 the thermal distribution $f_{thermal}$ applies only  for the transverse
 momenta, is very strong. One expects that during the 
evolution of the system the local momentum distribution is naturally 
driven towards an isotropic thermal distribution $f_{thermal}(p)$ depending
 on the total momentum $p$ of
 the particle in the local fluid element rest frame.
An anisotropic momentum distribution is natural in the initial condition, 
reflecting the presence of the longitudinal direction in the collision.
 However, during the expansion of the system, the density drops.
 Local thermalization processes adjust the temperature,
 which decreases as well. Similar, but not necessarily the same, 
processes would also lead  to the isotropization of the momentum distribution.
Hydrodynamics requires local equilibrium, with only small
 deviations introduced in the form of viscous corrections. 
 The range of deviations from equilibrium we consider,
 i.e. the relaxation
 from a system with two-dimensional equilibrium, 
which is far from the three-dimensional one, to the  usual isotropic 
equilibrium, breaks the condition that the corrections are small.
In fact in the initial state the correction 
to the longitudinal pressure is equal in magnitude and opposite in sign
 to the equilibrium pressure itself. It means
 that the generalization we propose must be understood as phenomenological 
description of the process of local equilibration.

The energy momentum tensor in the local rest fame is allowed to have an
anisotropic
 pressure. It starts with two components of nonzero pressure in the 
transverse direction and relaxes to a full isotropic one. 
It is written as a sum of an ideal fluid energy momentum-tensor 
and a  correction
\begin{equation}
T^{\mu\nu}=\left(\begin{array}{cccc}
\epsilon & 0 & 0 & 0 \\
0& p & 0 & 0 \\
0 & 0 & p & 0 \\
0 & 0 & 0 & p
\end{array}\right)+
\left(\begin{array}{cccc}
0& 0 & 0 & 0 \\
0& \Pi/2 & 0 & 0 \\
0 & 0 & \Pi/2 & 0 \\
0 & 0 & 0 & -\Pi
\end{array}\right)
\label{eq:rftensor}
\end{equation}
Maximal asymmetry at the initial time  means $\Pi(\tau_0)=p(\tau_0)$. 
For the subsequent evolution of the deviation from the equilibrium pressure
we assume a relaxation equation similar to Eq. (\ref{eq:dynamical})
\begin{equation}
\Pi(\tau)=\Pi(\tau_0)\exp(-(\tau-\tau_0)/\tau_\pi) \ ,
\label{eq:relax}
\end{equation}
which means that apart from the initial deviation from equilibrium 
shear viscosity effects are small.
 The simplest 
relaxation equation (\ref{eq:relax}) is the most natural assumption 
for a phenomenological description of the local equilibration.
 Microscopic processes behind the isotropization of the pressure could be 
collisions or instabilities \cite{Mrowczynski:2006de}. Without a reliable 
estimate of the relaxation timescale we use 
a phenomenological constant parameter $\tau_\pi$ . 

For the boost-invariant   flow the hydrodynamic equations reduce to
 the evolution of the energy density 
\begin{equation}
\frac{d \epsilon(\tau)}{d \tau}=-\frac{\epsilon(\tau) + p(\tau)}{\tau}+ 
\Pi(\tau_0)\exp(-(\tau-\tau_0)/\tau_\pi) \ . 
\label{eq:disshy}
\end{equation}
The above  equation is equivalent to an entropy  
production equation
\begin{equation}
\frac{d \left(s(\tau)  \tau\right)}{d \tau}= \frac{\Pi(\tau)}{T} \ . 
\label{eq:entropytpi}
\end{equation}
 Deviations from the 
ideal energy-momentum tensor lead to a gradual entropy production 
\cite{IS,Elze:2001ss}.
If we relate the entropy per unit rapidity to the particle multiplicity,
 a constraint on the dissipative effects in the hydrodynamic evolution 
appears \cite{Dumitru:2007qr}.
If the ratio of the entropies  at the end at at the beginning of 
the hydrodynamic evolution was known, one could estimate 
its production in the dissipative hydrodynamics.
A measure of the ratio of the final to initial entropy is given as the 
 ratio of the 
particle multiplicity per unit rapidity as observed in heavy-ion 
collisions to the multiplicity predicted in models without a hydrodynamic 
collective stage. It is difficult to falsify 
or confirm directly the predictions of such models, 
since a strongly interacting collective phase in the dynamics, 
such as the hydrodynamic evolution, cannot be excluded. An  argument
 in favor of such models
is that  they predict the centrality dependence of the 
multiplicity \cite{Kharzeev:2004if,Bialas:2006kw}. These models 
explain through an initial state effect,
the 
increased multiplicity of particles produced 
in heavy-ion collisions as compared
to proton-proton collisions and its centrality dependence at the same time.
 Consequently, there is not much additional 
entropy production allowed during any supplementary  hydrodynamic evolution. 
However, a different scenario is not excluded. 
Assuming that the initial entropy (multiplicity) 
per participating nucleon is the  same as in nucleon-nucleon
collision, the increased particle production  in heavy-ion collision
 is due to the entropy production during the hydrodynamic 
evolution and the hadron cascade stages.
 Nontrivial  centrality dependence of the multiplicity of produced particles 
can be explained  within a core-mantle model \cite{Bozek:2005eu}. 
The interaction region is composed of a dense core, 
where hydrodynamic evolution and entropy production 
takes place and  an outer mantle, where after initial particle production
 not much rescattering (entropy production) takes place. 
Depending on the centrality, the ratio of the core and 
mantle volume changes, reproducing the centrality dependence of the multiplicity scaled by the number of participating nucleons \cite{Bozek:2005eu}.
The relative increase of the multiplicity in the core region is of 60\%. 
This number represents an upper limit on allowed entropy production 
in the collective stage of the 
evolution of the fireball. For the most central collisions 
95\% of the particles are emitted from the dense core 
 \cite{Bozek:2005eu}. 

\section{One-dimensional expansion and entropy production}
 \label{sec:entropy}

\begin{figure}
\includegraphics[width=.8\textwidth]{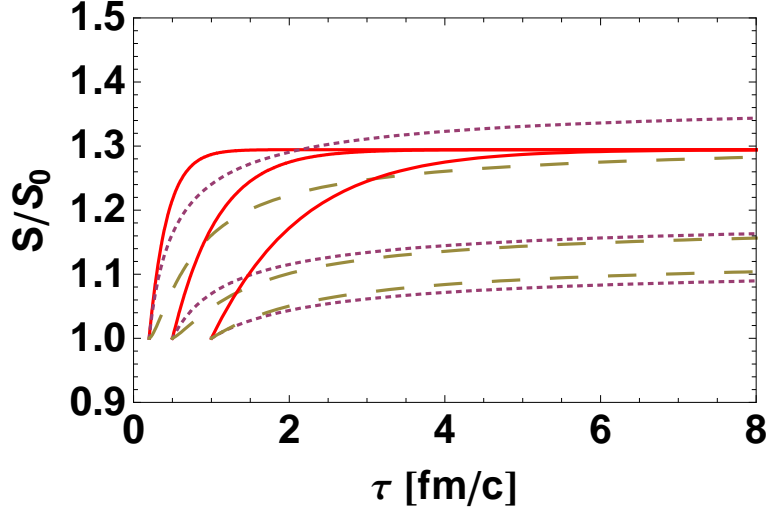}
\caption{Relative increase of the entropy from dissipative processes in the early stage of the collision 
for several initial times $\tau_0$ of the evolution. 
The dotted line represents the entropy production 
from the  Navier-Stokes shear viscosity tensor (\ref{eq:ns})
with $\eta=0.1\ s$, the  dashed 
line represents the increase of the entropy 
obtained from the second order viscous 
 hydrodynamic equation (\ref{eq:dynamical})
 with   $\eta=0.1 \ s$, $\tau_\pi=6 \eta/T s$,
 and $\Pi(\tau_0)=\frac{4\eta}{3\tau_0}$, and the solid 
represents the relative entropy production due to the 
stress tensor term of the form 
$\Pi(\tau)=p(\tau_0)\exp(-(\tau-\tau_0)/\tau_0)$ (\ref{eq:entropytpi}).
\label{fig:entropy}}
\end{figure}

For the one-dimensional Bjorken expansion and the relativistic 
gas equation of state $p=\epsilon/3$,
 the solution of the dissipative hydrodynamic equation
(\ref{eq:disshy}) can be written in a scaling form
\begin{eqnarray}
\epsilon(\tau)&=& f(\tau/\tau_0,\tau_\pi/\tau_0) \nonumber \\
& =& \epsilon(\tau_0)\left[e^\xi \xi\left(u^{2/3}E_{2/3}(1/\xi) -u
E_{2/3}(u/\xi)\right)\right. \nonumber \\
& & \left. +(9+3\xi)u^{2/3}-3e^{1/\xi}e^{u/\xi}\right]/(9u^2)
\end{eqnarray}
with $u=\tau/\tau_0$, $\xi=\tau_\pi/\tau_0$, and $E_x(z)=\int_1^\infty 
\frac{e^{-z/t}}{t^x} dt$. Similar 
scaling forms can be written for the pressure or the entropy.
At the initial time we have $\frac{d p(\tau)}{d\tau}|_{\tau_0}=-\frac{p(\tau_0)}
{\tau_0}$ and  $\frac{d \Pi(\tau)}{d\tau}|_{\tau_0}=-\frac{\Pi(\tau_0)}
{\tau_\pi}=-\frac{p(\tau_0)}{\tau_\pi}$. From the requirement $\Pi(\tau)\le 
p(\tau)$ we get $\tau_\pi\le\tau_0$. Within our phenomenological 
generalization of  hydrodynamics we obtain a bound on the relaxation 
time $\tau_\pi$.

\begin{figure}
\includegraphics[width=.8\textwidth]{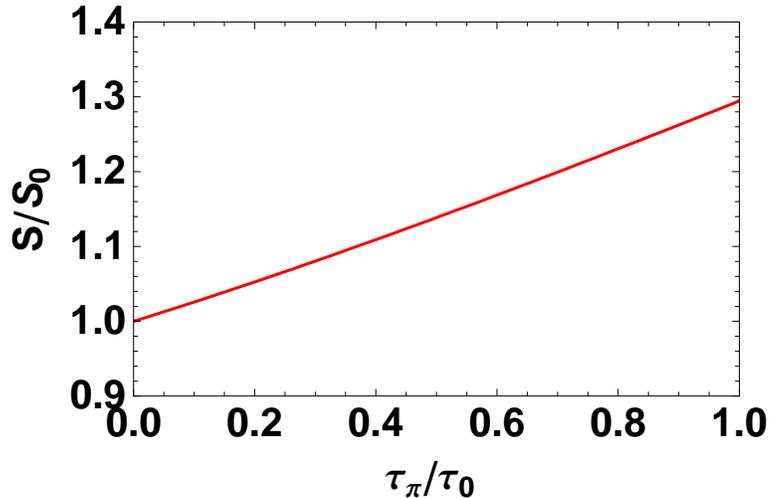}
\caption{Relative increase of the entropy from
 dissipative processes in the early stage of the collision 
as a function of the scaled relaxation time of the pressure anisotropy $\tau_\pi/\tau_0$.
\label{fig:entropytpi}}
\end{figure}

For the scaling solution the relative 
entropy production is a function of $\tau_\pi/\tau_0$ only 
\begin{equation}
\lim_{\tau\rightarrow \infty} \frac{s(\tau) \tau}{s(\tau_0)\tau_0}=
\lim_{\tau\rightarrow \infty} \frac{\tau}{\tau_0}(f(\tau/\tau_0
,\tau_\pi/\tau_0))^{3/4} \ .
\end{equation}
In Fig. \ref{fig:entropy} the entropy increase is 
plotted for several values of the initial starting times 
$\tau_0=0.2,\ 0.5, 1$fm/c and for $\tau_\pi/\tau_0=1$.
Independently of the starting time of the hydrodynamic
evolution the relative entropy production is the same  $28.4\%$.
 The entropy production is limited to the time of the order of $4 \tau_\pi$
 after $\tau_0$, 
whereas in the case of nonzero shear viscosity the dissipative processes are
 taking place through the whole evolution. (dashed and dotted lines in 
Fig. \ref{fig:entropy}).
The entropy production in the dissipative expansion depends on the ratio
 of the relaxation time $\tau_\pi$ and the initial time $\tau_0$ 
(Fig. \ref{fig:entropytpi}).
In the following we take $\tau_\pi=\tau_0$ exhibiting  the largest effects 
of the dissipative phase.

\section{Radial expansion}

\begin{figure}
\includegraphics[width=.8\textwidth]{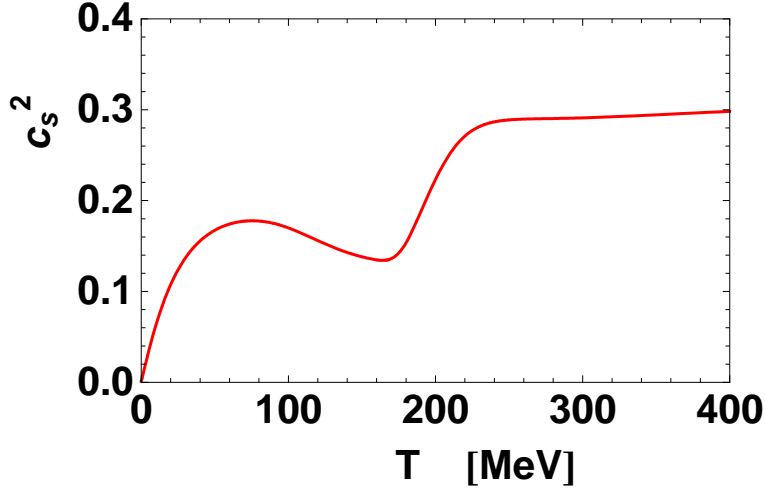}
\caption{Square of the velocity of sound as a function of the temperature
 for an equation of state interpolating between the  hadron gas and the 
 quark-gluon plasma
\cite{Chojnacki:2007jc}.
\label{fig:cs}}
\end{figure}

We consider a system with azimuthal symmetry  and boost-invariance  in the 
longitudinal direction. The initial conditions are given by no transverse flow 
and a Bjorken flow in the longitudinal direction. The initial 
entropy profile in the transverse
 plane (as a function of the transverse radius  $r$) 
is given by the wounded nucleon distribution  from the Glauber model
for central collisions,
with the nucleon-nucleon  cross section of $42$mb \cite{Miller:2007ri}.
 In the 
local rest frame the energy-momentum tensor is given by 
(\ref{eq:rftensor}). 
 The correction to the pressure $\Pi(\tau,r)$ 
describing  the pressure anisotropy in the local frame  is again given by
$\Pi(\tau,r)=p(\tau_0,r)\exp\left(-(\tau-\tau_0)/\tau_\pi\right)$.
The energy momentum tensor is boosted by the transverse and 
longitudinal velocity of the fluid element, and the 
 equation \cite{Baier:2006um}
\begin{equation}
u^\gamma \partial_\gamma \epsilon= -(\epsilon+p)
 \nabla_\mu u^\mu +\frac{1}{2} \Pi^{\mu\nu} 
 < \nabla_\mu u_\nu >
\end{equation}
and the radial component of
\begin{equation}
(\epsilon+p) u^\gamma \partial_\gamma u^\mu= \nabla^\mu p 
-\Delta^\mu_\nu \nabla_\alpha 
\Pi^{\nu\alpha}+\Pi^{\mu\nu}u^\gamma \partial_\gamma u_\nu
\end{equation}
are solved numerically 
for the transverse velocity and the energy density. The equation 
of state used 
is a combination of the lattice results at large temperatures and a
massive hadron gas equation of state at lower temperatures and was presented by
 Chojnacki and Florkowski
 (Fig. \ref{fig:cs}). The details of the parameterization are given in Ref.
\cite{Chojnacki:2007jc}, the equation of state exhibits only
 a moderate softening around the critical temperature $T_c=170$MeV.
The dissipation happens  early, in the plasma phase.
We consider two different starting times $\tau_0=0.5$ and $1$fm/c 
and for each case both the ideal and dissipative hydrodynamic evolutions.
The initial temperature at the center of the fireball is $300$MeV and $365$MeV
for the initial time of $1$fm/c and $0.5$fm/c respectively,
 for the ideal hydrodynamics. For the dissipative evolution 
we scale the initial entropy, corresponding to the chosen
 equation of state by a factor $1/1.3$
 to accommodate for the entropy production in the hydrodynamics phase. 
Such a procedure leads to similar total multiplicities after the
freeze-out in all the simulations.

\begin{figure}
\includegraphics[width=.5\textwidth]{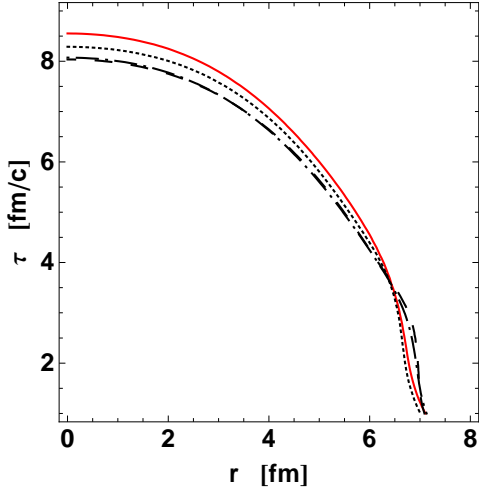}
\caption{Freeze-out hypersurfaces in the radial direction 
for $T_f=165$MeV. Solid and dashed-dotted line are for the ideal hydrodynamics
starting at $\tau_0=1$fm/c and $\tau_0=0.5$fm/c respectively.The 
dotted and dashed lines are  for the dissipative evolution corresponding to
$\tau_0=1$fm/c and $\tau_0=0.5$fm/c.
\label{fig:hyper}}
\end{figure}

The freeze-out hypersurfaces at  the temperature $T_f=160$MeV 
in  transverse-direction-time plane are shown 
in Fig. \ref{fig:hyper}. The effect of the slower cooling in
 the longitudinal direction in the evolution with dissipation
 is compensated by the  reduction of the initial temperature. As a result, the 
time to reach the freeze-out temperature is very similar for in the scenarios.

We calculate the transverse momentum spectra assuming boost invariance and 
 Boltzmann distributions for both pions and protons.
Corrections to the local equilibrium are significant only 
in the very early stage of the collision. For  the dominant part 
of the freeze-out hypersurface dissipative corrections to the 
statistical distribution function are negligible.  From the
 Cooper-Frye formula \cite{Cooper:1974mv}
one obtains in this case \cite{Rischke:1996em}
\begin{eqnarray}
\frac{d^3N}{d^2p_\perp dy}&=&
\frac{1}{2\pi^2}\int_0^{r_{max}} r dr \tau(r)
[ m_\perp K_1(\gamma_\perp m_\perp/T)
I_0(v_\perp\gamma_\perp p_\perp/T)  \nonumber \\
& -&  \frac{d\tau(r)}{dr}p_\perp K_0(\gamma_\perp m_\perp/T)
I_1(v_\perp\gamma_\perp p_\perp/T) ] \ ,
\label{eq:spectra}
\end{eqnarray}
where $m_\perp=\sqrt{p_\perp^2+m^2}$, $v_\perp$ is the transverse velocity and
$\gamma_\perp=1/\sqrt{1-v_\perp^2}$, $\tau(r)$ parameterizes the 
freeze-out hypersurface in the range of transverse radii $[0,  \  r_{max}]$, 
and $K_i$, $I_i$ are the Bessel functions.
\begin{figure}
\includegraphics[width=.7\textwidth]{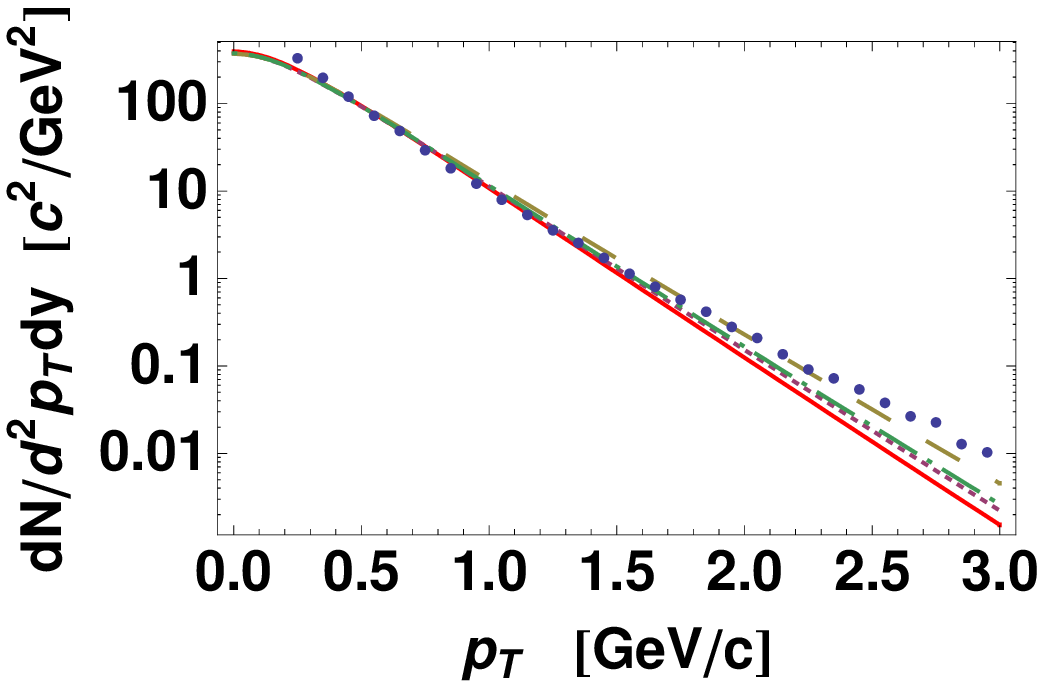}
\caption{$\pi^+$ spectra from  hydrodynamic calculations 
(same lines as in figure \ref{fig:hyper}). Data are from
 the PHENIX Collaboration \cite{Adler:2003cb} for most central events (0-5\%).
\label{fig:pip}}
\end{figure}
\begin{figure}
\includegraphics[width=.7\textwidth]{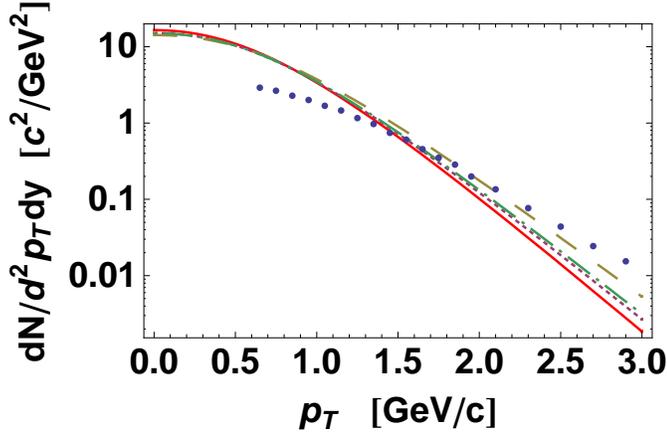}
\caption{Proton spectra from  hydrodynamic calculations 
(same lines  as in figure \ref{fig:hyper}). Data are from the
 PHENIX Collaboration \cite{Adler:2003cb} for most central events (0-5\%).
\label{fig:pro}}
\end{figure}
At the temperature of $165$MeV most of the 
particles are resonances, which only latter decay
 into pions and protons. The decays modify mainly the low momentum
 part of the spectra. We take this 
 effect and the nonzero baryon chemical potential 
 $\mu\simeq30$MeV into account by multiplying 
the obtained direct pion and proton spectra (Eq. \ref{eq:spectra})
 by a factor $4$ for pions and
 $5 \times \exp^{\mu/T}$ for protons  \cite{Torrieri:2004zz}. The slope
 of the high momentum part of the spectra is not modified 
considerably by resonance decays.

The dissipative stage in the hydrodynamic evolution leads to an increased
 transverse pressure, this drives a stronger transverse flow, and gives 
flatter spectra (larger effective slopes) for the same freeze-out temperature.
Such a stronger build-up of the transverse flow for viscous
 hydrodynamics has been observed \cite{Song:2007fn}. Combining a
small initial time for the hydrodynamic evolution with a dissipative 
increase of the transverse pressure one can qualitatively reproduce the 
observed effective slopes in transverse momentum spectra 
(dashed lines in Figs. \ref{fig:pip} and \ref{fig:pro}). More
 detailed analysis should include resonance decays and effects 
of hadronic rescattering, combined with a study of the elliptic flow.
 The increase in the transverse flow from the early dissipation 
is similar for both initial 
starting times of the evolution; the integrated dissipative 
effects are comparable,
 as observed already for the relative entropy increase
 (Sec. \ref{sec:entropy}).

\begin{figure}
\includegraphics[width=.85\textwidth]{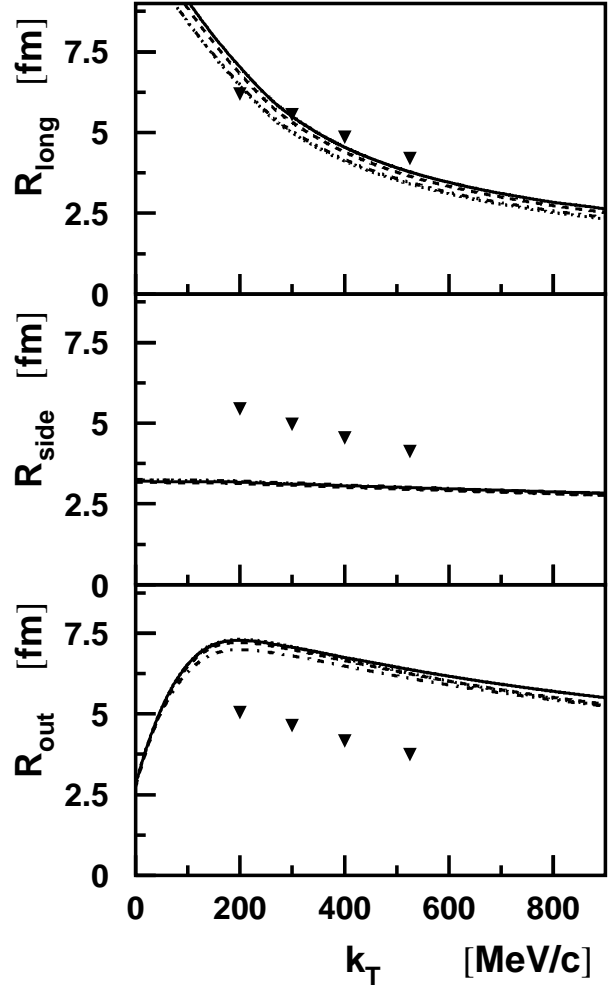}
\caption{HBT radii from hydrodynamic calculations as a function of the 
average particle momentum in the pair
 (same lines as in figure \ref{fig:hyper}). Data are from the STAR
 Collaboration \cite{Adams:2004yc} for most central events (0-5\%).
\label{fig:hbt}}
\end{figure}

The emission of indistinguishable particle pairs  from 
the fireball causes  two-particle quantum correlations.
 In the Bertsch-Pratt formula for the 
two-particle correlation function $C(p_1,p_2)$,
 particle 
momenta $p_1$ and $p_2$ in the longitudinally comoving frame
 of the pair 
 are parameterized by~: 
the average momentum $k_\perp$ and the three components of 
the relative momentum of the pair, 
$q_{long }$ along the longitudinal axis, $q_{out}$ along $k_\perp$,
 and $q_{side}$ in the third  perpendicular direction 
\cite{Bertsch:1989vn,Pratt:1986ev}.
 Using a Gaussian formula in the three directions
 $C(p_1,p_2)=1+\exp(-R_{long}^2q_{long}^2-R_{out}^2q_{out}^2-R_{side}^2q_{side}^2)$,
 the HBT radii $R_{long}$, $R_{out}$,
 and $R_{side}$ can be extracted from the width 
of the correlation function at mid-height. For that purpose, 
the two-particle correlation  function  on the freeze-out
hypersurface is calculated for three kinematic 
configurations
$p_{1,2}=k\pm q_{long}/2$, $p_{1,2}=k\pm q_{out}/2$, 
and $p_{1,2}=k\pm q_{side}/2$.
Explicit formulas are given in Ref. \cite{Rischke:1996em}.
In Fig. \ref{fig:hbt} the HBT radii are plotted as a function of the
 average transverse  momentum of the particles in the pair.
The effect of the early dissipation on the HBT radii is negligible, all
 the calculations lead to essentially the same HBT radii. This is in
 contrast to a strong sensitivity of the HBT radii on the viscous 
effects {\it at the freeze-out} as noticed in Refs. 
\cite{Teaney:2003kp,Romatschke:2007jx}. The HBT radii obtained 
in our hydrodynamic evolution cannot reproduce the observed ratio 
 $R_{out}/R_{side}$. The agreement with the experiment 
could be improved following the  recent work 
\cite{Chojnacki:2007rq},
 combing modified initial conditions, including a more
 complete hadron spectrum at freeze-out 
and the decay of resonances. The effects of the early dissipation
 could be included in such realistic calculations, and would require 
a retuning of the initial temperature to accommodate the additional entropy
 production in the early phase.
A different question is related to the effects of nonzero shear or 
bulk viscosities. Nonzero 
viscosities  modify the latter stages of the evolution and
lead to  significant modifications of the final  
obseravbles\cite{Romatschke:2007jx,Teaney:2003kp,Song:2007fn,Kharzeev:2007wb,Torrieri:2007fb}. This additional dissipation can be taken into 
account besides the early dissipation discussed in the present work.

\section{Discussion}

We propose a dissipative relaxation mechanism to
describe the relaxation of the pressure tensor from
 a two-dimensional initial state to the isotropic three-dimensional form. 
The final effect of this early dissipative stage depends on the value of the
relaxation time. Reduced work in the longitudinal direction
 and increased transverse pressure lead to a faster build-up of the 
transverse flow in the very early stage.
A relative entropy increase of up to 30\% is possible. Also a hardening 
of the transverse momentum spectra of particles at the freeze-out is noticeable.
Since at the freeze-out the corrections to the ideal 
energy-momentum tensor disappear,
we find very little effect on the HBT radii. After the system 
relaxes to the isotropic equilibrium state its evolution could be
 described by ideal fluid hydrodynamics with an initial transverse 
flow \cite{Chojnacki:2004ec}. We find that the effect of the
 asymmetry of the pressure and its subsequent 
relaxation could be significant for relaxation times of the order of $1$fm/c.
We  analyze the isolated effect of the initial dissipation only.
The initial dissipation described in this paper could 
be accompanied by a standard shear viscosity with a non-negligible 
viscosity coefficient $\eta$ an a different relaxation time $\tau_\pi$ 
in the latter evolution.

\section*{Acknowledgments}
The author is grateful to Miko\l aj Chojnacki and Wojtek 
 Florkowski for valuable discussions and making available 
the parameterization of the equation of state. The work has 
been supported in part by Polish Ministry of Science and 
Higher Education under grant N202 034 32/0919
\bibliography{hydr}

\end{document}